\documentclass[hidelinks, 11pt]{article}

\usepackage[utf8]{inputenc}
\usepackage[english]{babel}
\usepackage{blindtext}
\usepackage[left=2cm, right=2cm, top=2cm, bottom=2cm]{geometry} 
\usepackage{amsmath,amsfonts}
\usepackage{amssymb}
\usepackage{algorithm}
\usepackage{algcompatible}
\usepackage{algpseudocode}
\usepackage{pifont} 
\usepackage{natbib}
\usepackage{setspace}
\usepackage{graphicx}
\usepackage{subcaption}
\usepackage{hyperref}
\usepackage{mathrsfs}
\usepackage{amsthm}
\usepackage{tikz}
\usepackage{cancel}
\usepackage{lipsum}
\usepackage{booktabs}
\usepackage{multicol}
\usepackage{multirow}
\usepackage{siunitx}
\usepackage{rotating}
\usepackage{titling}

\DeclareTextAccent{\myacc}{T1}{4}

\newcommand{\commentsymbol}{//}
\algrenewcommand\algorithmiccomment[1]{\hfill \commentsymbol{} #1}

\newtheorem{claim}{Claim}[]

\allowdisplaybreaks

\def\keywordname{{\bfseries \emph Keywords}}%
\def\keywords#1{\par\addvspace\medskipamount{\rightskip=0pt plus1cm
\def\and{\ifhmode\unskip\nobreak\fi\ $\cdot$
}\noindent\keywordname\enspace\ignorespaces#1\par}}

\theoremstyle{definition}
\newtheorem{definition}{Definition} [section]

\title{Identifying Neural Connectivity using Bernoulli Autoregressive Partially Linear Additive Models}
\author{Carla Pinkney$^{1,^*}$, Carolina Eu\'{a}n$^1$ \& Alex Gibberd$^1$ \vspace{0.3cm} \\
    \small{$^1$ STOR-i Centre for Doctoral Training, School of Mathematical Sciences, Lancaster University, LA1 4YR, UK}\\ 
\footnotesize{$^*$ Correspondence to: c.pinkney@lancaster.ac.uk} \vspace{0.2cm}
\\}

\begin{document}
\maketitle

\begin{abstract}
       Characterising the interactions between spiking neurons is central to our understanding of cognitive processes such as memory, perception and decision making.
   In this work, we consider the problem of inferring connectivity in the brain network from non-stationary high-dimensional spike train data. Under a binned spike count representation of these data, we propose a Bernoulli autoregressive partially linear additive (BAPLA) model to identify the effective connectivity of a population of neurons. Estimates of the model parameters are obtained using a regularised maximum likelihood estimator, where an $\ell_1$ penalty is used to find sparse and interpretable estimates of neuronal interactions. We also account for non-stationary firing rates by adding a non-parametric trend to the model and provide an inference procedure to quantify the uncertainty associated with our estimated networks of neuronal interactions.  We use synthetic data to assess the performance of the BAPLA method, highlighting its ability to detect both excitatory and inhibitory interactions in various settings. Finally, we apply our method to a neural spiking dataset from the DANDI archive, where we study the interactions of neural processes in reaction to various stimulus-response type neuroscience experiments.   
\end{abstract}

\section{Introduction}
Characterising the synchronisation of spiking neurons from recorded signals is a challenging and active research problem in neuroscience \citep{keeley2020modeling, ren2020model,vareberg2024inference}. In the past decade, there have been significant advances in neural recording technologies, allowing for the exciting opportunity to study large scale neural activity in the living brain. For example, the first generation of Neuropixels probes enabled the simultaneous recording of hundreds of neurons distributed across the mouse brain, during a single experimental trial \citep{jun2017fully}. More recent advances include the development of the Neuropixels 2.0 and Neuropixels Ultra probes, which allow for recordings from even larger neuronal populations in small brains, during behaviour and free movement \citep{steinmetz2021neuropixels,ye2024ultra}. 
The use of high-throughput neural probes result in recordings of noisy, high-dimensional data, which present interesting challenges for both neuroscientists and statisticians. In this paper, we seek to understand how populations of neurons interact, and present a statistical methodology which can be used to estimate networks of neural interactions. The development of such a tool, which is capable of extracting accurate and interpretable descriptions of interactions from multi-region neural recordings, is imperative to our continued understanding of the brain's processes. 

In neuroscience, a common and widely accepted measure of functional interactions between neurons is that of cross-correlation \citep{narayanan2009methods}. 
Traditionally, neuroscientists have studied neural correlations using a variant of the cross-correlation histogram \citep{perkel1967neuronal} including, but not limited to, the joint peri-stimulus time histogram (JPSTH) \citep{gerstein1972mutual}, the snowflake plot \citep{perkel1975nerve, czanner2005theory} or the shuffle-corrected cross-correlogram \citep{aertsen1989dynamics, brody1999correlations}. These methods, which are readily available in popular neuroscience data analysis packages such as Pynapple and SpikeInterface, are regularly used for the analysis of electrophysiological data. However, these methods are limited in the sense that they only study a handful of neurons at a time, and are therefore ignore likely contributions from other neurons. Consequently, detected correlations could be ambiguous as a result of the possible, and yet excluded, influence of other neurons \citep{zhao20121}.

Generalised linear models (GLMs) are also routinely used by neuroscientists for the analysis of simultaneously recorded neural data. Popularised by \cite{brillinger1988maximum}, who studied small networks of 3 neurons, GLMs offer a highly interpretable framework for the analysis of firing rates, and have also been used for a variety of other problems in computational neuroscience including encoding \citep{paninski2004maximum, truccolo2005point}, decoding \citep{gao2003quantitative} and for the assessment of neural interactions \citep{zhao20121}.

\begin{figure}[t]
\centering
\begin{subfigure}[b]{0.49\textwidth}
    \includegraphics[width =0.9\textwidth]{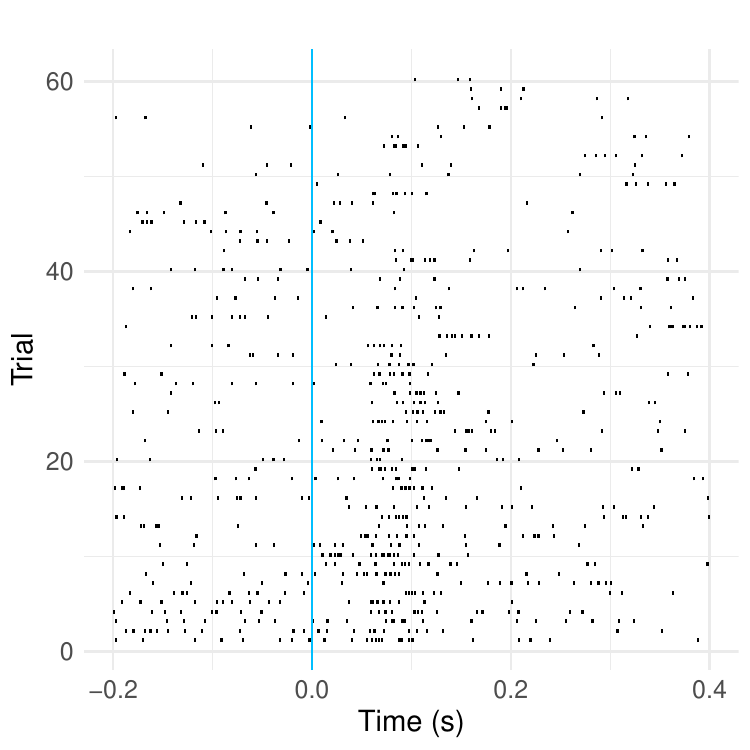}
    \caption{}
    \end{subfigure}
    \hfill
    \begin{subfigure}[b]{0.49\textwidth}
    \includegraphics[width =0.9\textwidth]{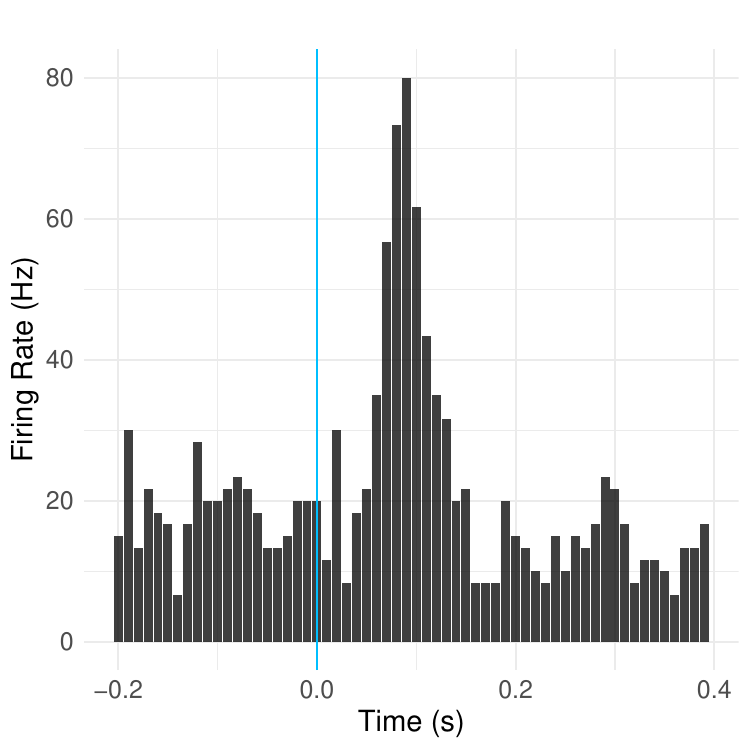}
    \caption{}
    \end{subfigure}
\caption{\small{Example neuron from the spike train dataset of \cite{steinmetz2019distributed}: (a) Raster plot showing single neuron response to 60 repeats of a visual stimulus and (b) peri-stimulus time histogram (PSTH) for the same data.}}
 \label{fig:raster_psth}
\end{figure}

In this paper, we extend the popular and widely used GLM framework in order to account for non-stationary dynamics that are often observed in stimulus-response type neuroscience experiments.Take, as example, the spike train data from \cite{steinmetz2019distributed} shown in Figure \ref{fig:raster_psth}. In this experiment, neural activity was recorded over a series of trials, while mice performed a visual discrimination task (see section 5 for full details).  Figure \ref{fig:raster_psth} shows a raster plot and Peristimulus time histogram (PSTH) aligned to the onset of a visual stimulus, at time $t=0$, for an example neuron in the dataset. Each row in the raster plot is representative of a single spike train, which represents the neural response, or more specifically the firing times of the neuron in a particular trial. By contrast, the PSTH describes the trial-averaged activity of the neuron, by pooling data across trials. It is clear to the see that the firing rate of the neuron increases in response to the visual stimulus, thus motivating the need for methods which account for inherent non-stationarities commonly observed in neural data.   It is in these stimulus-response type experiments where the methodology outlined in this paper would be of most use. 

Often, neural spike trains are represented  in continuous time as a multivariate point process $\mathbf{N}(t)=\{N_i(t)\}_{i\in \{1, \dots, d\}}$ whose $i^{th}$ component gives the number of spikes observed in neuron $i$ in the time interval $(0,t], t\leq n$.
In the statistics literature, the multivariate Hawkes \citep{hawkes1971point} process has received significant attention, and has become a popular tool for the identification and modelling of functional connectivity from recorded neural spike trains \citep{lambert2018reconstructing, wang2021joint, wang2024statistical}. Conversely, 
in the neuroscience literature, neural spike trains are often represented in terms of their binned spike count, which refers to a discretised version of the continuous time process \citep{kass2011assessment}. The so-called binned spike count is constructed by decomposing the time axis into bins of equal width $\delta$, resulting in a discrete integer-valued time series where each value indicates the number of action potentials (spikes) within a specified time bin, i.e., $Y_{t,i}=N_i(t+\delta) - N_i(t), t\leq n,$ for each $i=1,\dots, d.$

Here, we take the latter approach, and select $\delta$ to be small enough to ensure that at most one spike per neuron can occur within a time bin, e.g., $\delta = 1$ms. Therefore, we will consider a binary time series representation of the neural spiking data, which takes value $1$ in a time bin where there is a spiking event, and $0$ everywhere else. 
Importantly, it can be shown that under the usual regularity condition, where no two events occur at the same time, that the likelihood function of the binary time series approximates the likelihood of the point process as $\delta \rightarrow 0$ \citep{kass2011assessment}.

In this work, we build on the regularised logistic model proposed in \cite{zhao20121} for the detection of neuronal interactions. There, the authors studied a network with $d=30$ neurons, 
and constructed a model assuming that neural firing rates remained stationary over time. 
In our approach, we account for non-stationary firing rates by adding a non-parametric component to the logistic model, and consider larger networks with more neurons. 
Admittedly, it is difficult to determine a `ground truth' for network interactions among populations of real neurons. 
Therefore, in addition to calibrating our estimator using synthetic data, we also outline an inference procedure that can be used to quantify the uncertainty associated with our estimated networks. 
To the best of our knowledge, our approach is the first to: 1) account for non stationary firing rates in the regularised logistic model for the estimation of effective connectivity between neurons, and 2) provide confidence intervals for these estimates.

The rest of this paper is organised as follows. In Section 2, we describe the methodology used for the detection of neural interactions, including the required estimation procedure, before giving details of our inference procedure in Section 3. In Section 4, we conduct a simulation study to demonstrate the satisfactory performance of our estimator. In Section 5, we apply our model to electrophysiology data obtained from the DANDI archive, where we study the interactions of neural processes in reaction to the stimulus-response type neuroscience experiments described in \cite{steinmetz2019distributed}. We conclude with a discussion in Section 6.

\section{Bernoulli Autoregressive Partially Linear Additive Models}

Consider a multivariate binary time series $\mathbf{Y}_t = (Y_{t,1}, \dots, Y_{t,d})^\top$ such that the series at time $t$ are conditionally independent of one another given past realisations, i.e., $$\Pr(\mathbf{Y}_t=\mathbf{y}_t|\mathbf{Y}_{<t}) =  \prod_{i=1}^d \Pr(Y_{t, i}=y_{t,i}|\mathbf{Y}_{<t}),$$
where $\mathbf{Y}_{<t}$ denotes the entire history of the multivariate series observed up to, but not including, time $t$.
\begin{definition}\citep{shojaie2022granger}.
    Time series $Y_j$ is Granger non-causal for time series $Y_i$ if and only if $\forall \ t$,
    \begin{equation}
        \Pr(Y_{t,i} = 1|Y_{<t,1}, \dots Y_{<t,j}, \dots, Y_{<t,d}) = \Pr(Y_{t,i}=1| Y_{<t,1}, \dots, Y_{<t,(j-1)}, Y_{<t,(j+1)}, \dots, Y_{<t,d}).
        \label{def:GC}
    \end{equation}
\end{definition}

Granger causality has proven to be an effective method for the analysis of neural spike train data, due to its ability to detect directional interactions amongst populations of neurons \citep{kim2011granger, lintas2018granger,casile2021robust}. We wish to develop a class of models that allow for easy elicitation of such Granger causal relations.

\subsection{Model Formulation}
Consider a set of spike trains, with non-stationary firing rates, recorded from a population of neurons. Under a binned spike count representation of these data, we propose to use the following Bernoulli autoregressive partially linear additive (BAPLA) model to identify the effective connectivity of this neural population. 

Let $\{Y_{t,i}\}_{t=1}^n \in \{0,1\}$ be the $i^{th}$ variate of the $d$-variate random vector $\mathbf{Y}_t$ for $1\leq i \leq d$, i.e., $Y_{t,i}$ denotes the state of neuron $i$ at time $t$. Then, the BAPLA model is defined as
 \begin{equation}
    Y_{t,i}|\mathbf{Y}_{t-1} \sim \operatorname{Bernoulli}\left(\frac{1}{1+\exp\{-\beta_{i}-\boldsymbol{\gamma}_i^\top\mathbf{Y}_{t-1}-f_i(t/n)\}}\right),
    \label{eq:bernoulli}
\end{equation}
where $\mathbf{Y}_{t-1} = (Y_{t-1,1}, Y_{t-1,2},\dots, Y_{t-1,d})^\top$ 
 denotes the previous values of the multivariate process, $\beta_{i}$ is a constant offset parameter, $\boldsymbol{\gamma}_i^\top \in \mathbb{R}^{1\times d}$ is an unknown parameter vector, and $f_i(t/n):[0,1]\rightarrow \mathbb{R}$ is an unknown smooth one-dimensional and differentiable function. To ensure identifiability of the intercept term $\beta_{i}$ and the smooth function $f_i(t/n)$, we impose the constraint that $\int_0^1 f_i(u) du=0$ for $i=1, \dots, d.$

We are interested in using the BAPLA model to investigate Granger casual interactions in high-dimensional non-stationary multivariate binary time series.  In the context of our neuronal analysis, the components of the BAPLA model have the following interpretations: 1) $\beta_{i}$ models the base-line firing rate of neuron $i$; 2) $\boldsymbol{\gamma}_i^\top $ provides an insight into the interactions between the spike train recorded from neuron $i$ and the spike trains from the whole neural population; and 3) $f_i(t/n)$ models the non-stationary firing rate of neuron $i$.
Admittedly, our proposed method can be considered as an extension to a particular case of the generalised linear autoregressive (GLAR) model previously studied in \cite{hall2018learning}. However, a key methodological advancement of the BAPLA model is its ability to capture non-linear dynamics via the time varying function $f_i(t/n)$. The inclusion of such a component is essential in order to account for the non-stationary dynamics commonly observed in real world data (Figure \ref{fig:raster_psth}).

The model in \eqref{eq:bernoulli} can be written in terms of the conditional distribution of observing a spike at time $t$ 
\begin{equation}
    \Pr(Y_{t,i}=1|\mathbf{Y}_{t-1}=\boldsymbol{y}_{t-1}) = \operatorname{logit}^{-1}\bigg(\beta_{i}  +\boldsymbol{\gamma}_i^\top \boldsymbol{y}_{t-1}+ f_i(t/n)\bigg),
    \label{prop_glm}
\end{equation}
where $\operatorname{logit}^{-1}(a)=1/1+\exp(-a)$. 
Using this representation, equation \eqref{prop_glm} can be used to infer a network of interactions $\boldsymbol{\Gamma} \in \mathbb{R}^{d \times d}$ which is constructed row-wise from $\boldsymbol{\gamma}_i^\top$ (i.e. $\boldsymbol{\gamma}_i^\top$ is the $i^{th}$ row of $\boldsymbol{\Gamma}$).
Therefore, given observed data $\{y_{t,i}\}_{t=1}^n$, for $i = 1,\dots, d$, one can identify and model connectivity via an estimate of the matrix $\boldsymbol{\Gamma}.$  In particular, by leveraging the component-wise structure of \eqref{prop_glm}, we can infer 
that time series $Y_i$ does not Granger cause series $Y_j$ if and only if ${\Gamma}_{ij}=0$. 

In general, it is reasonable to assume that interactions among a population of real neurons are sparse. For example, a cortical neuron receives between $10^3-10^4$ inputs, which is a particularly small fraction of the $10^{10}$ neurons in the cerebral cortex \citep{zhao20121}. Consequently, we now introduce the following notation for the true, and assumed sparse, interaction matrix $\boldsymbol{\Gamma}$.  Firstly, we define the edge set $ E(\boldsymbol{\Gamma}):=\{(i,j):\Gamma_{ij}\neq 0, i \neq j\},$ and let $\operatorname{card}(E(\boldsymbol{\Gamma})) \leq s$. That is, the set of off-diagonal entries in $\boldsymbol{\Gamma}$ which determine the edge set of the graph, is at most $s$. Going forward, we will refer to the true matrix $\boldsymbol{\Gamma}$ as being $s$-sparse, i.e., the corresponding graph has at most $s$ edges. In addition, we denote by $\tilde{d} = \max_{i = 1,\dots,d}|\{j \in \{1, \dots, d\}: \Gamma_{ij}\neq 0\}|,$
 the maximum degree of a node, which stipulates the maximum number of non-zero entries in any row of $\boldsymbol{\Gamma}$. Finally, we will assume that entries in the interaction matrix are bounded, i.e., $\boldsymbol{\Gamma}\in [\gamma_{\min}, \gamma_{\max}]^{d \times d}.$

It is also necessary to place certain assumptions on the time varying function $f_i(t/n)$. Firstly, we assume that each $f_i(t/n)$ satisfies some smoothness conditions, such as being differentiable up to a certain order. 
Taking advantage of this smoothness, we approximate each $f_i(t/n)$ as a weighted summation of a set of basis functions
\begin{equation*}
    f_i(t/n) \approx \sum_{k=1}^m c_{ik}\phi_{ik}(t/n) \quad \textrm{for} \ t = 1,2,\dots,n,
\end{equation*}
where $\phi_{ik}(\cdot)$ and $c_{ik}$ are the basis functions and their coefficients respectively. Consequently, we can write a parametric form of model \eqref{prop_glm} as
\begin{equation*}
    \Pr(Y_{t,i}=1|\mathbf{Y}_{t-1}=\boldsymbol{y}_{t-1}) = \operatorname{logit}^{-1}\left(\beta_{i}  +\boldsymbol{\gamma}_i^\top \boldsymbol{y}_{t-1}+ \sum_{k=1}^m c_{ik}\phi_{ik}(t/n)\right).
    \label{prop_glm2}
\end{equation*}

\subsection{Parameter Estimation and Model Selection}

In our estimation procedure for the BAPLA model, we model the non-stationary firing rate $f_i(t/n)$ of each neural spike train as a linear combination of known basis functions. While there are a wide variety of basis functions that could be used, we choose the B-spline basis due to its simplicity and numerical properties.
Generally, the number of B-spline basis functions $m$ is chosen to be large enough to ensure that the fitted curve is representative of the data. In our estimation procedure, we specify a fixed number of basis functions over a given sequence of knots, chosen to be equally spaced for $t \in [0,n]$. 
Specifically, let $\boldsymbol{\phi}_i(t/n) = (\phi_{i,1}(t/n), \dots, \phi_{i,m}(t/n))$ be the associated B-spline basis for each $i = 1,\dots, d$. Then, we estimate 
$\hat{f}_i(t/n) = \mathbf{\hat{c}}_i^\top\boldsymbol{\phi}(t/n)$
where $\mathbf{\hat{c}}_i$ is the estimated vector of coefficients for each neuron $i$.

An obvious extension to our model would be the inclusion of an additional penalty term to automatically select the number of basis functions, rather than specifying a fixed $m$. However, we emphasise that the main goal of this work is to get an accurate estimate of the interaction matrix $\boldsymbol{\Gamma}$, for which a reasonably good estimate of $f_i(t/n)$ is sufficient. Indeed, we will show later on in the simulation studies that there exists an $a>0 \in \mathbb{Z}^+$ such that for $m>a$, the estimation error of the interaction matrix $\boldsymbol{\Gamma}$ does not improve. Thus, selecting a `large enough' $m$ is sufficient for the purposes of this work.

We obtain estimates of the model parameters using a \textbf{regularised maximum likelihood estimator}. Given data $\{y_{t,i}\}_{t=1}^n $ for $i =1,\dots, d$, we estimate $\hat{\theta} = (\hat{\boldsymbol{\gamma}}_i, \hat{\beta}_{i}, \hat{c}_{i,1},\dots, \hat{c}_{i,m})$ by maximising the following penalised log-likelihood
\begin{align}
L(\theta,\boldsymbol{\phi}_i(t/n),\lambda) :=& \bigg{[}  \sum_{t=1}^n y_{t,i} \bigg{(}\beta_{i}+ \sum_{k=1}^m c_{i,k} \phi_{i,k}(t/n) + \boldsymbol{\gamma}_i^\top \boldsymbol{y}_{t-1}\bigg{)} \nonumber\\
    &  - \log\bigg{(}1+\exp\bigg{\{}\beta_{i}+ \sum_{k=1}^m c_{i,k} \phi_{i,k}(t/n)+\boldsymbol{\gamma}_i^\top \boldsymbol{y}_{t-1}\bigg{\}}\bigg{)} - \lambda \|\boldsymbol{\gamma}_i\|_1 \bigg{]},
   \label{obj_c}
\end{align}
where $\|\boldsymbol{\gamma}_i\|_1 = \sum_{j=1}^d |\Gamma_{ij}|$ denotes the $\ell_1$ norm and $\lambda>0$ is a regularisation parameter, such that
\begin{align}
    \hat{\theta}:=\arg\max_{\boldsymbol{\theta}}  &\quad L(\theta,\boldsymbol{\phi}_{i}(t/n),\lambda) \nonumber  \\
   \textrm{such that}& \quad \sum_{k=1}^mc_{ik}\phi_{i,k}(t/n) = 0 \quad \textrm{for} \ t=1,\dots, n\;.
   \label{constraint}
\end{align}

The above regularised maximum likelihood estimator attempts to find a sparse estimate of the interaction matrix $\boldsymbol{\Gamma}$ whilst also ensuring a good fit to the data by modelling the non-linear interactions via $f_i(t/n)$.  
The constraint in \eqref{constraint} is the sample analogue of the assumption that 
$\int f_i(u)du=0$, which ensures identifiability of the intercept term $\beta_{i}$ and the smooth non-linear function $f_i(t/n)$. 

Following the work of \cite{guo2013variable}, we convert \eqref{obj_c} and \eqref{constraint} into an unconstrained problem by centering the basis functions as follows. Let 
\begin{equation*}
    \bar{\phi}_{i,k}(t/n) = \frac{1}{n}\sum_{t=1}^n\phi_{i,k}(t/n), \quad \textrm{and} \quad \tilde{\phi}_{i,k}(t/n) = \phi_{i,k}(t/n) - \bar{\phi}_{i,k}(t/n).
\end{equation*}
Then, estimates of the model parameters $\theta = (\hat{\boldsymbol{\gamma}}_i, \hat{\beta}_{i}, \hat{c}_{i,1},\dots,\hat{c}_{i,m} )$ are obtained by maximising the unconstrained problem, i.e.,
\begin{equation}
\hat{\theta} = \arg\max_{\boldsymbol{\theta}} L(\theta,\tilde{\boldsymbol{\phi}}_{i}(t/n),\lambda)\;,\label{obj_unc}
\end{equation}
and the corresponding estimator of the unknown smooth function is 
$\hat{f}_i(t)=\sum_{k=1}^m \hat{c}_{i,k}\tilde{{\phi}}_{i,k}(t/n)$. Note, for simplicity, in the rest of the paper we will assume that $\phi_{i,k}(t/n)=\phi_{k}(t/n)$, i.e. the basis functions are chosen to be the same for all neurons $i=1,\ldots,d$.

\subsection{Computational Choices}

The estimator in \eqref{obj_unc} requires careful specification of a tuning parameter $\lambda$. 
Several approaches have been considered in the literature, including the Akaike information criterion (AIC), the Bayesian information criterion (BIC), and cross-validation procedures. 
In this work, the regularisation parameter is selected using a BIC-type criterion due to its computational efficiency over cross-validation procedures which require additional model fittings. Moreover, BIC tends to select larger values of $\lambda$ compared to AIC, thereby yielding a more interpretable representation of effective connectivities between neurons. 

Using a training group of simulated data, we obtain estimates of $\boldsymbol{\gamma}_i$ over a fixed search grid of $\lambda$ values and choose the tuning parameter which minimises
$$\mathrm{BIC}_\lambda(\hat{\boldsymbol{\gamma}}_i) =  -2 \ell (\hat{\boldsymbol{\gamma}}_i) +  \log(n)\times (k + 1 + m),$$
where $\ell(\boldsymbol{\hat{\gamma}}_i)$ denotes the maximised  log-likelihood of the associated model and $k$ denotes the number of non-zero entries in $\boldsymbol{\hat{\gamma}}_i$. These values are then averaged across the number of samples in the training set to arrive at a final optimal parameter $\lambda^*$.

We solve the optimisation problem in \eqref{obj_unc} 
 using a \textbf{coordinate descent algorithm} \citep{friedman2010regularization}.
However, consider first the  \textit{unpenalised} equivalent of \eqref{obj_unc} which can be solved via the Newton algorithm, which amounts to iteratively re-weighted least sqaures \citep{friedman2023glmnet}.
That is, if we have current estimates of the parameters $(\tilde{\beta}_{i}, \tilde{\boldsymbol{\gamma}}_i, \tilde{c}_{i,1}, \dots, \tilde{c}_{i,m})$, then we can use a Taylor expansion around these estimates to obtain a quadratic approximation to the log-likelihood

\begin{algorithm}[t]
\small
\raggedright
\caption{Estimation Procedure}
\begin{algorithmic} 
\State
\State
\For{$i=1,\dots, d$} 
\State Initialise $\lambda$ and ${\tilde{\boldsymbol{\theta}}}_0$
\State Sample basis functions 
\State Compute $\ell_Q(\tilde{\beta}_{i}, \tilde{\boldsymbol{\gamma}}_i, \tilde{c}_{i1}, \dots, \tilde{c}_{im})$
\State Use cyclic coordinate descent to solve the penalised weighted least-squares problem
    \begin{equation*}
        \min_{\boldsymbol{\theta}}\{-\ell_Q(\tilde{\beta}_{i},\tilde{\boldsymbol{\gamma}}_i,\tilde{c}_{i,1}, \dots, \tilde{c}_{i,m})+\lambda \|\boldsymbol{\gamma}_i\|_1\}.
    \label{eq:pen_ls}
    \end{equation*}
    \State \hspace{0.5cm} Calculate the coordindate-wise update step for each element in $\boldsymbol{\gamma}_i^\top$, i.e., for $j=1,\dots, p$, 
        \begin{equation}
            \tilde{\gamma}_{i,j} \leftarrow \frac{\mathcal{S}\left(\sum_{t=1}^n w_{t,i}y_{j, t-1}\left(z_{t,i} - \tilde{y}^{(j)}_t\right), \lambda\right)}{\sum_{t=1}^n w_{t,i}(y_{j, t-1})^2},
            \label{eq:coord_update}
        \end{equation}
       \State \hspace{0.5cm} where $\tilde{y}_t^{(j)}=\tilde{\beta}_0 + \boldsymbol{\tilde{\gamma}}_i^\top \boldsymbol{y}_{t-1} - \tilde{\gamma}_{i,j} y_{j,t-1} +  \sum_{k=1}^m \tilde{c}_{i,k} \tilde{\phi}_k(t/n)$ is the fitted value excluding the
       \State \hspace{0.5cm} contribution from $y_{j, t-1}$ and $\mathcal{S}(a,\kappa)$ is the soft-thresholding operator with value
        \begin{align*}
           \mathcal{S}(a,\kappa) = \begin{cases}
                a - \kappa \quad &\textrm{if} \ a >0 \ \textrm{and} \ \kappa<|a| \\
                a+\kappa  \quad &\textrm{if}  \ a<0 \ \textrm{and} \ \kappa<|a| \\
                 0  \quad &\textrm{if} \ \kappa\geq|a|. 
            \end{cases}
        \end{align*}
    \State \hspace{0.5cm} Update the intercept term
    \begin{equation}
        \tilde{\beta}_{i} \leftarrow \frac{\sum_{t=1}^n w_{t,i}(z_{t,i}- \sum_{k=1}^m \tilde{c}_{i,k} \tilde{\phi}_k(t/n) - \boldsymbol{\gamma}_i^\top \boldsymbol{y}_{t-1})}{\sum_{t=1}^n w_{t,i}}
        \label{eq:intercept}
    \end{equation}

    \State \hspace{0.5cm} Update the spline coefficients, i.e. for $k=1,\dots,m$
    \begin{equation}
        \tilde{c}_{i,k} \leftarrow \frac{\sum_{t=1}^n w_{t,i}\phi_k(t/n)(z_{t,i}-\tilde{\beta}_{i}-\tilde{\gamma}_i^\top \boldsymbol{y}_{t-1}-\sum_{j\neq k}\tilde{c}_{i,j}\tilde{\phi}_j(t/n))}{\sum_{t=1}^n w_{t,i}\tilde{\phi}^2_k(t/n)}
        \label{eq:spl}
    \end{equation}
    \State  \hspace{0.5cm} Repeat steps $(7)-(9)$ until convergence.
\State
\EndFor
\label{algorithm}
\end{algorithmic}
\end{algorithm}

\begin{align*}
    \ell_Q(\beta_{i}, \boldsymbol{\gamma}_i, c_{i,1}, \dots, c_{i,m})=-\frac{1}{2}&\sum_{t=1}^n \bigg{\{}w_{t,i}\bigg{(}z_{t,i}-\beta_{i}- \sum_{k=1}^m c_{i,k} \tilde{\phi}_k(t/n)-\boldsymbol{\gamma}_i^\top \boldsymbol{y}_{t-1}\bigg{)}^2\bigg{\}} \\&+ C\bigg{(}\tilde{\beta}_{i}, \tilde{\boldsymbol{\gamma}}_i, \tilde{c}_{i1}, \dots, \tilde{c}_{im}\bigg{)}^2,
\end{align*}
where $C(\cdot)$ is a constant with respect to the new parameter positions, and we have
\begin{align*}
    z_{t,i} &= \tilde{\beta}_{i} + \tilde{\boldsymbol{\gamma}}_i^\top \boldsymbol{y}_{t-1} +  \sum_{k=1}^m \tilde{c}_{i,k} \tilde{\phi}_k(t/n) + \frac{y_{t,i}-\tilde{p}(\boldsymbol{y}_{t-1})}{w_{t,i}},  &\textrm{(working response)}\\ 
    w_{t,i} &= \tilde{p}(\boldsymbol{y}_{t-1})(1-\tilde{p}(\boldsymbol{y}_{t-1})),  &\textrm{(weights)} 
\end{align*}
where $\tilde{p}(\boldsymbol{y}_{t-1})$ is evaluated at the current parameters, i.e., \\
$\tilde{p}(\boldsymbol{y}_{t-1})= {1}/{1+ \exp\left\{-\tilde{\beta}_{i}- \sum_{k=1}^m \tilde{c}_{i,k} \tilde{\phi}_k(t/n)-\tilde{\boldsymbol{\gamma}}_i^\top \boldsymbol{y}_{t-1}\right\}}.$ The Newton update is then obtained by minimising $\ell_Q.$ 
The approach we take to maximise the regularised estimator \eqref{obj_unc} is very similar. We compute the quadratic approximation $\ell_Q$ about the current parameters, and then we use coordinate descent to solve the penalised weighted least squares problem 
\begin{equation*}
    \min_{\boldsymbol{\theta}}\{-\ell_Q(\beta_{i},\boldsymbol{\gamma}_i, c_{i,1}, \dots, c_{i,m})+\lambda \|\boldsymbol{\gamma}_i\|_1\}.
\end{equation*}
Simple calculus yields the coordinate-wise update steps for each of the model parameters outlined in Algorithm 1.

\section{Inference}
\label{sec:inference}
In general, it is difficult to determine a `ground truth' for network interactions amongst a population of real neurons. As noted in \cite{zhao20121}, \textit{``...even if microscopic examination of living tissue were feasible, it would still be unclear what the strength and sign of a synaptic interaction between two cells might be"}. It is therefore fundamental to recognise the uncertainty associated with our estimates of the interaction matrix. However,  despite the vast progress made in recent years for the estimation of brain connectivity from neural recordings, little work has been done to assess the uncertainty associated with these estimates. In this section, we outline an inference procedure for the construction of confidence intervals for entries in $\boldsymbol{\Gamma}$, with the overarching aim of quantifying the uncertainty around our estimates of the interactions between neural processes. 

\subsection{Desparsifying the Regularised Estimator}

The construction of confidence intervals for the interaction matrix is not a straightforward task. Due to the lasso regularisation in \eqref{obj_unc}, entries in $\hat{\boldsymbol{\gamma}}_{i} \in \mathbb{R}^d$ can be exactly zero, leading to a discontinuous limiting distribution which may have probability one at zero whenever the true parameter value is zero \citep{waldorp2024network}. As such, existing techniques for calculating standard errors are not suitable \citep{tibshirani1996regression} and standard methods for bootstrapping also do not work \citep{fu2000asymptotics, chatterjee2011bootstrapping}.
To circumvent this problem, we will construct a desparsified estimator of the coefficient vector $\boldsymbol{\gamma}_i$, for all $i=1,\dots, d$. This technique was pioneered by \cite{van2014asymptotically} in the context of linear and generalised linear models, and laid the ground work for inference procedures in other high-dimensional models \citep[see for example][]{javanmard2014confidence, gueuning2016confidence}.

Consider the negative log-likelihood for a particular dimension $i$ corresponding to the $t^{th}$ sample observation 
\begin{equation}
\label{log_lik2}
     \ell(\beta_{i}, \boldsymbol{\gamma}_i, c_{i,1}, \dots, c_{i,m})= - \sum_{t=1}^n y_{t,i} \log\{p_i(t)\} + (1-y_{t,i})\log\{1-p_i(t)\},
\end{equation}
where $p_i(t) = [1+\exp\{-\beta_{i}-\boldsymbol{\gamma}_i^\top \boldsymbol{y}_{t-1} -f_i(t/n)\}]^{-1}$.
Following the work of \cite{van2014asymptotically}, we obtain the following expressions for the first and second partial derivatives (of the sample mean) of \eqref{log_lik2} with respect to ${\boldsymbol{\gamma}_i}\in \mathbb{R}^d$ 
\begin{align}
\nonumber
    \dot{\ell}_{\boldsymbol{\gamma}_i} = \frac{\partial}{\partial \boldsymbol{\gamma}_i}\ell(\beta_{i}, \boldsymbol{\gamma}_i, c_{i,1}, \dots, c_{i,m}) &= -\frac{1}{n} \sum_{t=1}^n \{y_{t,i}-p_i(t)\}\boldsymbol{y}_{t-1}, \\
    \ddot{\ell}_{\boldsymbol{\gamma}_i}= \frac{\partial}{\partial \boldsymbol{\gamma}_i \partial \boldsymbol{\gamma}_i^\top}\ell(\beta_{i}, \boldsymbol{\gamma}_i, c_{i,1}, \dots, c_{i,m})  &= \frac{1}{n} \sum_{t=1}^n p_i(t)\{1-p_i(t)\} \boldsymbol{y}^\top_{t-1}\boldsymbol{y}_{t-1}.
    \label{eq:FI}
\end{align}
Letting $\hat{\Sigma}:= \ddot{\ell}_{\hat{\boldsymbol{\gamma}}_i}$, we can then define the desparsified estimator of the coefficient vector $\boldsymbol{\gamma}_i$, for each $i=1,\dots, d$ as 
\begin{equation}
    \hat{\boldsymbol{\gamma}_i}^{desp} = \hat{\boldsymbol{\gamma}_i} - \hat{\Theta} \dot{\ell}_{\hat{\boldsymbol{\gamma}}_i},
    \label{eq:desp}
\end{equation}
where $\hat{\Theta}:= \hat{\Sigma}^{-1}$. 

Often, spike train data are considered to be high dimensional in the sense that the dimension $d$ (or equivalently the number of neurons) is large in comparison to the sample size $n$ \citep{pinkney2024regularised}. However, under a binned spike count representation of these data, it is likely that $d<n$ meaning that the matrix inverse $\hat{\Sigma}^{-1}$ can be easily obtained. Conversely, in scenarios where $d>n$ or even $d\approx n$ it would be necessary to construct a `relaxed' inverse $\hat{\Theta}$ due to singularity of $\hat{\Sigma}$. Indeed, a variety of methods have been proposed for this purpose, though the most widely used are that of nodewise regression \citep{van2014asymptotically} and convex optimisation \citep{javanmard2014confidence}

\subsection{Constructing Confidence Intervals}

Our desparsification approach is essentially the method proposed in \cite{van2014asymptotically} for generalised linear models, adapted to include an additional smooth component via $f_i(t/n)$. We therefore hypothesise that the term 
$\hat{\Theta} \dot{\ell}_{\hat{\boldsymbol{\gamma}}_i}$  will prevent the asymptotic distribution from containing a point-mass at zero, thus enabling the construction of confidence intervals for each of the entries in $\hat{\boldsymbol{\gamma}}_i$. While a full theoretical analysis for this modification to the desparsified estimator is left for future work, our empirical results (Section 4) provide preliminary support for the validity of our proposed approach.  
We now state the following claim which we will use to construct confidence intervals for entries in the interaction matrix. 
\begin{claim}
    Let $\hat{\boldsymbol{\gamma}}_i^{desp}$ denote the desparsified estimator defined in \eqref{eq:desp}. Under certain regularity conditions, we expect that
    \begin{equation*}
        \sqrt{n}(\hat{\boldsymbol{\gamma}}_i^{desp} - \boldsymbol{\gamma}_i)/\hat{\boldsymbol{\sigma}}_i = \boldsymbol{V}_i + {o}_p(1),
    \end{equation*}
    where $\boldsymbol{V}_i$ converges weakly to a $\mathcal{N}(0,1)$-distribution and $\hat{\boldsymbol{\sigma}}_i= \sqrt{\operatorname{diag}(\hat{\Theta}\hat{\Sigma}\hat{\Theta})}$.
\end{claim}
Provided that Claim 1 holds, an asymptotic $(1-\alpha)100\%$ confidence interval for  $\Gamma^*_i$ is given by $\hat{\boldsymbol{\gamma}}_i^{deb}\pm \Phi^{-1}(1-\alpha/2)\boldsymbol{\sigma}_i/\sqrt{n},$
where $\Phi(\cdot)$ denotes the standard normal cumulative distribution function. 

\section{Synthetic Experiments}
\subsection{Simulation Setting}
\label{sec:sim_set}
We use synthetic data to assess the accuracy of the BAPLA model in recovering estimates of the interaction matrix and non-stationary firing rates in various simulation settings. 
In the first instance, we simulate binary time series data according to model \eqref{eq:bernoulli} for multivariate processes of  dimension $d=10$ and $d =50$. We obtain  simulated data for $n=1000, 2000, 5000$ and $10,000$ time points, which is equivalent to recording periods (in length seconds (s)) of $1$s, $2$s, $5$s and $10$s, provided we select $\delta = 1$ms. 

\begin{figure}[t]
\centering
\begin{subfigure}[b]{0.32\textwidth}
    \includegraphics[width =\textwidth]{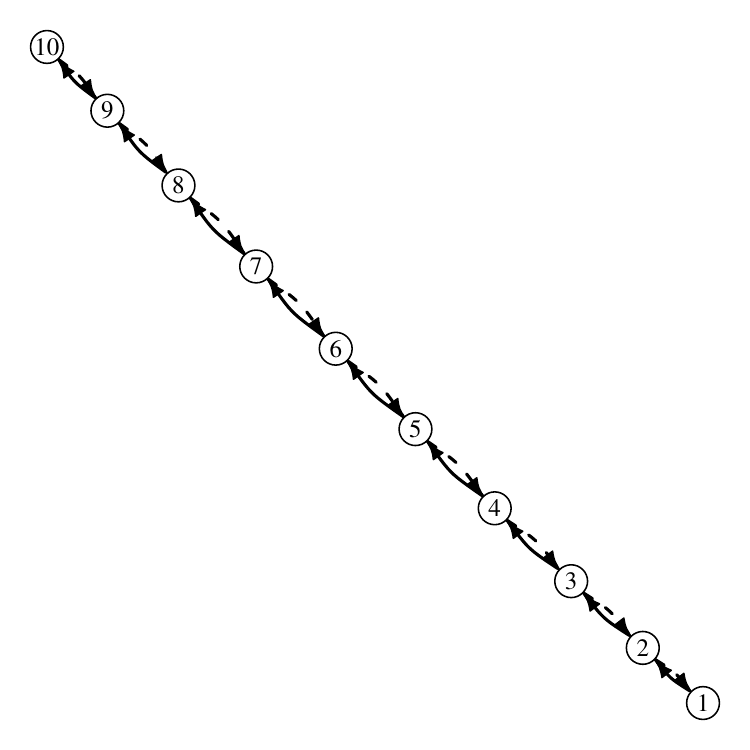}
    \caption{}
    \label{net_a}
    \end{subfigure}
    \hfill
    \begin{subfigure}[b]{0.32\textwidth}
    \includegraphics[width =\textwidth]{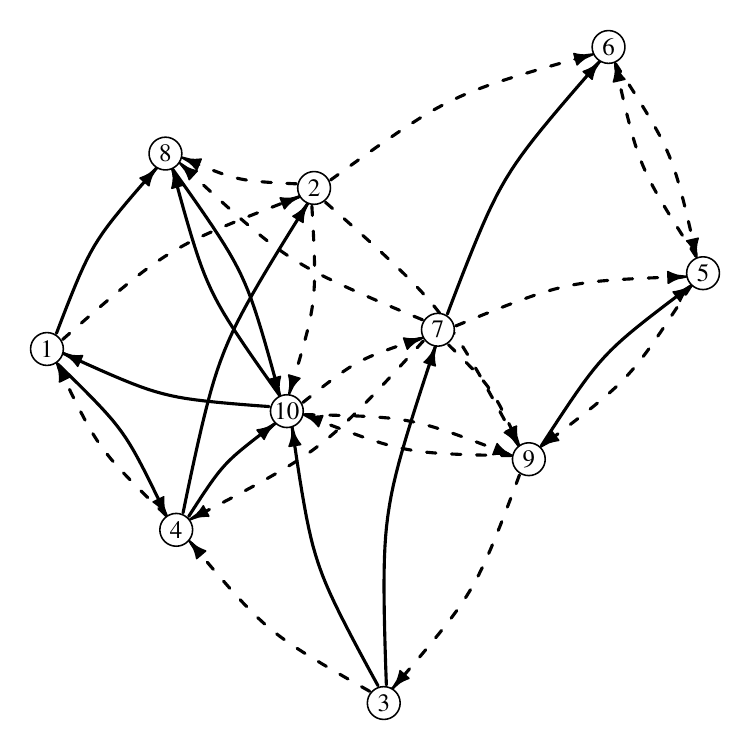}
    \caption{}
    \label{net_b}
    \end{subfigure}
    \begin{subfigure}[b]{0.32\textwidth}
    \includegraphics[width =\textwidth]{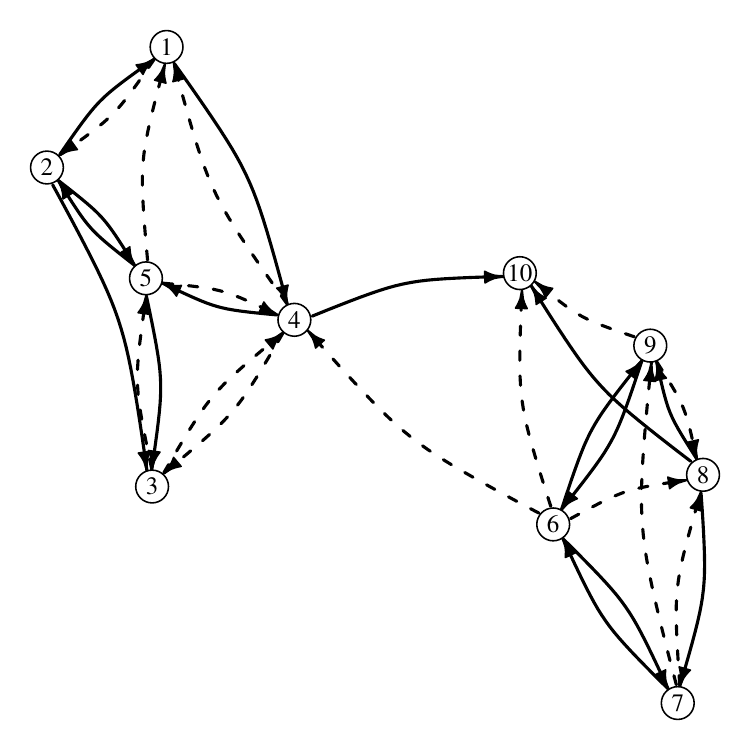}
    \caption{}
    \label{net_c}
    \end{subfigure}
\caption{\small{Illustrations of the different network structures used in the simulations with $d=10$ neurons. (a) Chain graph. (b) \text { Erdős-Rényi }graph. (c) Stochastic block model. Dashed edges are inhibitory interactions ($\Gamma_{ij}<0$) and solid edges are excitatory interactions ($\Gamma_{ij}>0$).}}
 \label{networks}
\end{figure}

We mainly focus on three types of network structure: a simple one in which only neighbouring neurons interact, a random one in which interactions are assigned arbitrarily, and a more complex one with connected sub-networks, in which edges are more common within a sub-network than between sub-networks. Mathematically, these are more commonly referred to as chain graphs (CG), Erdos-Renyi (ER) random graphs, and graphs generated via the stochastic block (SB) model---examples can be seen in Figs. \ref{net_a}, \ref{net_b}, and \ref{net_c} respectively. 

In our simulations, we generate realistic neural firing patterns via careful specification and parametrisation of the BAPLA model. Firstly, we allow for both excitatory and inhibitory interactions in the simulated networks, encoded by positive and negative entries in the interaction matrix, respectively. Secondly, to mimic the behaviour of neural spike trains observed in stimulus-response experiments, we set $f_i(t/n)$ in model \eqref{eq:bernoulli} as either the scaled probability density function (pdf) for a normal or gamma distribution, and vary this throughout our different simulation scenarios. 
Thirdly, we consider different specifications for the intercept term $\beta_{i}$, which determines the baseline firing rate of the neurons. In our population of real neurons, we observe firing rates in the range $10$Hz - $200$Hz (Figure \ref{fig:psth}). To approximate  this behaviour, we set $\beta_{i} = -2.6$, realising a firing rate of approximately $70$Hz, in which case we observe that approximately $16\%$ of the simulated data points are spiking events, i.e., $y_{t,i}=1$. We also consider a setting where  $\beta_{i}=0.1$. In doing so, we are able to assess the impact of different neural firing rates on our estimation procedure.  Going forward, we will refer to the settings $\beta_{i} =-2.6$ and $\beta_{i} = 0.1$ as low- and high-firing, respectively. 
Since our focus here is on the estimation of the interaction matrix, we specify networks with a similar number of edges, enabling a fair comparison between the different simulation settings. Moreover, each interaction matrix has both positive and negative entries of the same magnitude, $|\Gamma_{ij}|=0.3$.

The performance of the BAPLA model is assessed across a variety of scenarios including: varying the complexity of the network (CG, ER or SB), the type of interaction (excitation or inhibition), the dimensionality of the network, the size of the data set ($1$s, $2s$, $5$s or $10$s recording periods), the baseline firing rate and the specification of the non-stationary firing rate (normal or gamma).
For each combination of model parameters, we obtain parameter estimates over $N=100$ Monte Carlo simulations. We compare estimates of the interaction matrix to the ground truth using the relative mean squared error (RMSE) defined as $\operatorname{RMSE}(\hat{\Gamma}) = \|\hat{\boldsymbol{\Gamma}} - \boldsymbol{\Gamma}\|^2_F/ \|\boldsymbol{\Gamma}\|^2_F.$ Estimates of the intercept term and non-stationary firing rate are assessed using a mean squared error criterion, which we denote by MSE$_{\beta}$ and MSE$_f$, respectively. 
We also report results on the recovery of the sparsity pattern of the interaction matrix using the area under receiver operating characteristic curve (AUC). Numerical results averaged over 100 repetitions are presented in Table \ref{Sim1_new2}.

Confidence intervals for entries in the interaction matrix are constructed using the methods described in Section \ref{sec:inference}. 
We calculate the average coverage (Avgcov) and average length (Avglen) of the confidence intervals for entries in the interaction matrix, corresponding to variables in the active set $s$ or its compliment $s^c$, i.e., non-zero and zero coefficients respectively. 
Specifically, we report empirical versions of
\begin{align*}
    \operatorname{Avgcov}_s = \frac{1}{|s|}\sum_{j \in s} \mathbb{P}(\Gamma_{ij} \in \operatorname{CI}_j) \quad \textrm{and} \quad
    \operatorname{Avglen}_s = \frac{1}{|s|}\sum_{j \in s} \operatorname{length}(\operatorname{CI}_j),
\end{align*}
where $\operatorname{CI}_j$ denotes the confidence interval for the $j^{th}$ element of $\boldsymbol{\gamma}_i^\top$. Analogous definitions are given for the set $s^c.$ 

\subsection{Simulation Results}
\label{sec:sim_results}
\begin{table}[]
\scriptsize
    \centering  
  \begin{tabular}{ccccccccccc}
\hline
 \multirow[b]{2}{*}{$d$} & \multirow[b]{2}{*}{$\beta_0$} & \multirow[b]{2}{*}{$n$}  & \multicolumn{4}{c}{\text {Estimation}} & \multicolumn{4}{c}{\text {Inference}}    \\

         &  & & RMSE$_{\Gamma}$ & MSE$_{\beta_0}$ & MSE$_{f}$ & AUC & AvgCov$_{s}$ & AvgCov$_{s^c}$ & AvgLen$_{s}$ & AvgLen$_{s^c}$
\\ 
\hline
\\
\multicolumn{11}{c}{Chain Graph} \\
\\
  10 & 0.10 & 1000 & 0.21 & 0.03 & 0.59 & 0.82 & 0.94 & 0.95 & 0.57 & 0.57 \\ 
     &  & 2000 & 0.12 & 0.04 & 0.60 & 0.91 & 0.93 & 0.95 & 0.40 & 0.40 \\ 
    &  & 5000 & 0.05 & 0.05 & 0.60 & 0.96 & 0.93 & 0.95 & 0.26 & 0.26 \\ 
     &  & 10000 & 0.03 & 0.05 & 0.61 & 0.98 & 0.92 & 0.95 & 0.18 & 0.18 \\ 
   \\
     & -2.60 & 1000 & 0.86 & 0.51 & 5.37 & 0.56 & 0.95 & 0.96 & 1.90 & 1.90 \\ 
     &  & 2000 & 0.61 & 0.12 & 1.42 & 0.64 & 0.96 & 0.96 & 1.31 & 1.32 \\ 
     &  & 5000 & 0.27 & 0.04 & 0.28 & 0.71 & 0.95 & 0.95 & 0.82 & 0.82 \\ 
     &  & 10000 & 0.16 & 0.04 & 0.24 & 0.73 & 0.94 & 0.95 & 0.58 & 0.58 \\ 
   \\
    50 & 0.10 & 1000 & 0.32 & 0.02 & 0.56 & 0.76 & 0.94 & 0.95 & 0.61 & 0.61 \\ 
     &  & 2000 & 0.20 & 0.02 & 0.57 & 0.85 & 0.95 & 0.95 & 0.42 & 0.42 \\ 
     &  & 5000 & 0.09 & 0.04 & 0.59 & 0.98 & 0.94 & 0.95 & 0.27 & 0.27 \\ 
     &  & 10000 & 0.05 & 0.04 & 0.60 & 1.00 & 0.94 & 0.95 & 0.61 & 0.61 \\ 
\\
     & -2.60 & 1000 & 0.94 & 0.39 & 4.46 & 0.53 & 0.96 & 0.96 & 2.36 & 2.36 \\ 
     &  & 2000 & 0.84 & 0.06 & 0.76 & 0.58 & 0.96 & 0.96 & 1.49 & 1.49 \\ 
     &  & 5000 & 0.40 & 0.03 & 0.27 & 0.69 & 0.96 & 0.95 & 0.89 & 0.89 \\ 
     &  & 10000 & 0.27 & 0.02 & 0.22 & 0.72 & 0.95 & 0.95 & 0.61 & 0.62 \\ 
   \\
\multicolumn{11}{c}{Erdos-Renyi Random Graph} \\
\\
    10 & 0.10 & 1000 & 0.88 & 0.02 & 0.33 & 0.61 & 0.94 & 0.94 & 0.63 & 0.63 \\ 
     &  & 2000 & 0.81 & 0.02 & 0.16 & 0.68 & 0.92 & 0.94 & 0.44 & 0.44 \\ 
     &  & 5000 & 0.58 & 0.03 & 0.09 & 0.82 & 0.87 & 0.92 & 0.28 & 0.28 \\ 
     &  & 10000 & 0.53 & 0.04 & 0.07 & 0.87 & 0.84 & 0.87 & 0.20 & 0.20 \\ 
   \\
    & -2.60 & 1000 & 1.03 & 0.33 & 1.30 & 0.51 & 0.94 & 0.94 & 1.70 & 1.69 \\ 
     &  & 2000 & 0.97 & 0.07 & 0.52 & 0.53 & 0.94 & 0.94 & 1.17 & 1.17 \\ 
     &  & 5000 & 0.92 & 0.03 & 0.11 & 0.57 & 0.94 & 0.94 & 0.73 & 0.73 \\ 
     &  & 10000 & 0.85 & 0.02 & 0.08 & 0.63 & 0.92 & 0.93 & 0.51 & 0.51 \\ 
   \\
    50 & 0.10 & 1000 & 0.95 & 0.13 & 0.34 & 0.56 & 0.95 & 0.95 & 0.65 & 0.65 \\ 
     &  & 2000 & 0.87 & 0.12 & 0.23 & 0.65 & 0.95 & 0.95 & 0.45 & 0.45 \\ 
     &  & 5000 & 0.60 & 0.07 & 0.12 & 0.86 & 0.94 & 0.95 & 0.28 & 0.28 \\ 
     &  & 10000 & 0.45 & 0.05 & 0.08 & 0.92 & 0.94 & 0.95 & 0.20 & 0.20 \\ 
   \\
     & -2.60 & 1000 & 1.02 & 0.14 & 1.44 & 0.51 & 0.95 & 0.95 & 2.06 & 2.07 \\ 
     &  & 2000 & 1.00 & 0.04 & 0.35 & 0.51 & 0.95 & 0.95 & 1.32 & 1.33 \\ 
     &  & 5000 & 0.97 & 0.02 & 0.14 & 0.53 & 0.95 & 0.95 & 0.79 & 0.79 \\ 
     &  & 10000 & 0.96 & 0.02 & 0.10 & 0.55 & 0.95 & 0.95 & 0.55 & 0.55 \\ 
   \\
\multicolumn{11}{c}{Stochastic Block Model} \\
\\
    10 & 0.10 & 1000 & 0.85 & 0.10 & 0.47 & 0.63 & 0.94 & 0.94 & 0.61 & 0.63 \\ 
     &  & 2000 & 0.77 & 0.08 & 0.35 & 0.72 & 0.92 & 0.95 & 0.43 & 0.44 \\ 
     &  & 5000 & 0.59 & 0.06 & 0.31 & 0.83 & 0.90 & 0.94 & 0.27 & 0.28 \\ 
     &  & 10000 & 0.42 & 0.05 & 0.28 & 0.93 & 0.85 & 0.93 & 0.19 & 0.20 \\ 
   \\
     & -2.60 & 1000 & 1.00 & 0.26 & 2.01 & 0.54 & 0.94 & 0.96 & 1.69 & 2.42 \\ 
     &  & 2000 & 0.94 & 0.09 & 0.65 & 0.56 & 0.93 & 0.95 & 1.17 & 1.67 \\ 
     &  & 5000 & 0.90 & 0.04 & 0.20 & 0.58 & 0.93 & 0.95 & 0.74 & 1.05 \\ 
     &  & 10000 & 0.80 & 0.04 & 0.17 & 0.65 & 0.88 & 0.94 & 0.52 & 0.74 \\ 
   \\
    50 & 0.10 & 1000 & 0.93 & 0.10 & 0.45 & 0.58 & 0.95 & 0.95 & 0.66 & 0.68 \\ 
     &  & 2000 & 0.87 & 0.08 & 0.36 & 0.65 & 0.94 & 0.95 & 0.46 & 0.47 \\ 
     &  & 5000 & 0.65 & 0.07 & 0.30 & 0.83 & 0.94 & 0.95 & 0.29 & 0.30 \\ 
     &  & 10000 & 0.52 & 0.06 & 0.29 & 0.90 & 0.93 & 0.95 & 0.20 & 0.21 \\ 
   \\
     & -2.60 & 1000 & 1.03 & 0.27 & 2.27 & 0.51 & 0.95 & 0.95 & 2.08 & 2.56 \\ 
     &  & 2000 & 1.00 & 0.07 & 0.47 & 0.51 & 0.95 & 0.95 & 1.38 & 1.72 \\ 
     &  & 5000 & 0.97 & 0.03 & 0.20 & 0.53 & 0.95 & 0.95 & 0.84 & 1.06 \\ 
     &  & 10000 & 0.95 & 0.03 & 0.16 & 0.56 & 0.94 & 0.95 & 0.59 & 0.74 \\  
     \\
   \hline
   \end{tabular}
    \caption{\small{Simulation results averaged over 100 replications for estimating the parameters of the BAPLA model.}} 
  \label{Sim1_new2}
\end{table}

Examining the results over all simulation settings, it is clear to see that more data, i.e. longer recording periods, yields more favourable results in terms of both RMSE and the AUC scores for the interaction matrix. We also observe lower RMSE and higher AUC scores for the high-firing $(\beta_{i}=0.1)$ setting compared to that of the low-firing. Importantly, we note that the diagonal elements of the Fisher information \eqref{eq:FI} are maximised when $p_i(t) \approx \operatorname{logit}^{-1}(0.1)\approx 0.5$. Thus, the high-firing scenario is (by design) a more informative setting and therefore we should expect better estimates of the model parameters in this case. In general,  lower-dimensional settings yield more favourable results, as evidenced by the lower RMSEs in cases where $d=10$ compared to $d=50$. 

In terms of complexity of the network structure, our results show that (perhaps as expected) the BAPLA model performs best on the most simple network (CG) achieving an AUC score of $>0.76$ under the high-firing setting regardless of the recording period length and $>0.96$ when $n\geq 5000$. 
Comparatively, the AUC scores of the ER and SB models are $>0.82$ under the high-firing setting for $n\geq 5000$. However, for the low-firing setting, the AUC scores fall as low as $0.51$ even for the longest recording period, highlighting the difficulty observed in detecting neuronal interactions in these settings. 

Turing our attention now to the recovery of the intercept term and non-stationary firing rates, we see that in general, the MSE reduces as a function of the sample size $n$ as expected. Importantly, we note that in the CG setting $f_i(t/n)$ is proportional to a normal pdf, whereas in the ER setting it is proportional to a gamma pdf. In the SB setting, specifications of $f_i(t/n)$ are alternated from block to block. Examining the results across all scenarios, we see that ER network yields more favourable results in terms of MSE$_f$, compared to that of the CG and SB networks, suggesting that the BAPLA model better recovers the shape of the gamma pdf compared to that of the normal pdf. Interestingly, we highlight that the recovery of the non-stationary firing rate appears to be insensitive to the dimensionality of the problem. That is, MSE$_f$ is comparable across the low and high-dimensional settings. This is in contrast to estimates of the interaction matrix, which were largely more favourable in lower dimensional settings  compared to higher-dimensional settings.  

Results for the inference procedure with the de-sparsified lasso estimator are also presented in Table \ref{Sim1_new2}. In general, the estimator appears to perform better for the entries in $s^c$ compared to those in $s$. However, we would argue that both scenarios yield sufficiently adequate coverage results. As expected, the average length of the confidence intervals reduce as a function of $n$ and are generally shorter for the high-firing setting compared to that of the low-firing.

\subsection{Sensitivity Analysis}
\label{sen}

\begin{figure}[t]
\centering
\begin{subfigure}[b]{0.48\textwidth}
    \includegraphics[width =\textwidth]{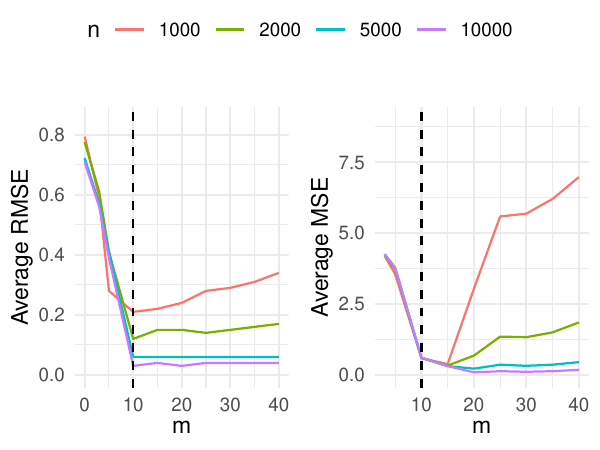}
    \caption{}
    \label{normal}
    \end{subfigure}
    \hfill
    \begin{subfigure}[b]{0.48\textwidth}
    \includegraphics[width =\textwidth]{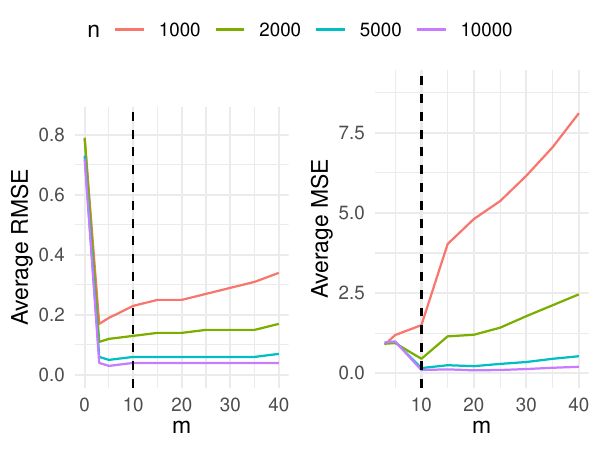}
    \caption{}
    \label{gamma}
    \end{subfigure}
\caption{\small{Simulation results averaged over 100 samples for estimating the interaction matrix (left) and the non-stationary firing rate (right) with a range of $m$ basis functions for (a) normal $f_i(t/n)$ and (b) gamma $f_i(t/n)$.}}
 \label{fig:varym}
\end{figure} 

In this section, we investigate the consequences of specifying a fixed number of basis functions $m$ for the estimation of the non-stationary firing rate $f_i(t/n)$ for each  $\ i=1,\dots, d$. Keeping in mind that our primary goal is to estimate the effective connectivity amongst a population of neurons, we show here that accurate estimation of the interaction matrix is in fact insensitive to the estimation of $f_i(t/n)$, provided that $m$ is sufficiently large (e.g., $m\geq10$). 
As an illustrative example, we mimic the simulation scenario in Section \ref{sec:sim_set} above, in which the interaction matrix is specified by a chain graph  (Fig \ref{net_a}) with dimension $d=10$. Data is simulated according to model \eqref{eq:bernoulli} with $\beta_{i} = 0.1$ and $f_i(t/n)$ is specified as either the scaled pdf of a normal or gamma distribution. As before, we generate data for $n = 1000, 2000, 5000$ and $10,000$ time points. 

Our goal is to investigate how sensitive estimation of the interaction matrix is to the number of basis functions used to estimate the non-stationary firing rate $f_i(t/n).$ Coupled with this, we are interested in exploring whether our choice of $m=10$ basis functions is sufficient for accurate estimation of $f_i(t/n)$.  To do so, we conduct a sensitivity analysis in which we fit the model multiple times for various values of $m$, i.e., $m\in\{3,5,10,15,20,25,30,35,40\},$ and evaluate various performance metrics associated with the parameter estimates of the BAPLA model. In doing so, our aim is to show that there exists some $a>0 \in \mathbb{Z}^+$ such that for $m>a$, the estimation error of the interaction matrix $\boldsymbol{\Gamma}$ does not considerably improve. Thus, selecting a `large enough' $m$ is sufficient for the purposes of this work. For completeness, we also investigate the case in which $m=0$ referring to model \eqref{eq:bernoulli} without including the $f_i(t/n)$ term, i.e., a standard logistic model. This scenario is included to motivate the need for an additional model component when there are inherent non-stationarities in the data.

\begin{figure}[t]
\centering
\begin{subfigure}[b]{0.32\textwidth}
    \includegraphics[width =\textwidth]{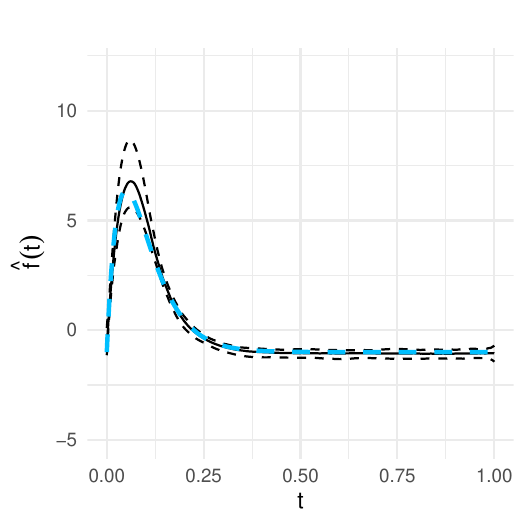}
    \caption{}
    \label{fig:con_a}
    \end{subfigure}
    \hfill
    \begin{subfigure}[b]{0.32\textwidth}
    \includegraphics[width =\textwidth]{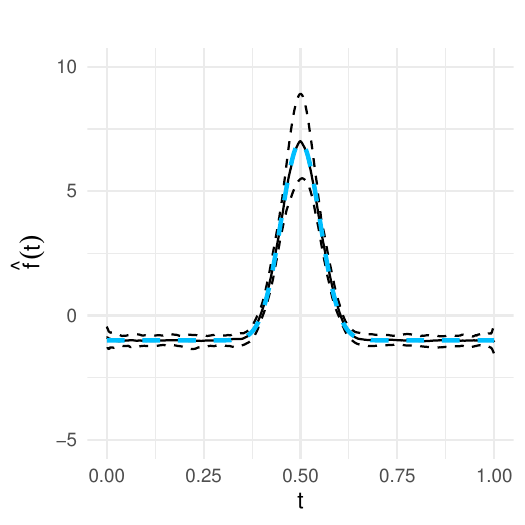}
    \caption{}
     \label{fig:con_b}
    \end{subfigure}
    \begin{subfigure}[b]{0.32\textwidth}
    \includegraphics[width =\textwidth]{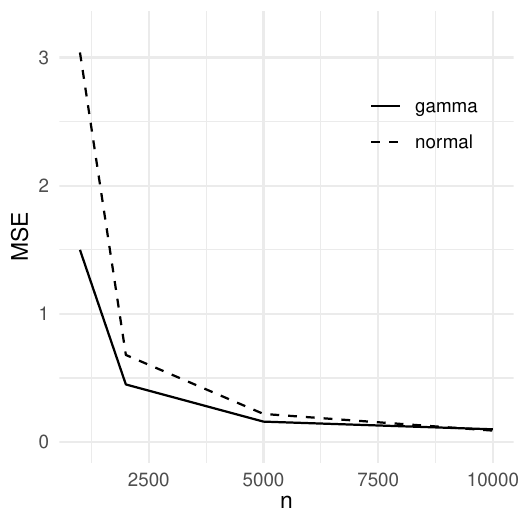}
    \caption{}
     \label{fig:con_c}
    \end{subfigure}
\caption{\small{Estimates of $f_i(t/n)$ for the (a) gamma and (b) normal model with $95\%$ confidence intervals (dashed lines) and compared to the ground truth (dashed blue line). Mean squared error as a function of $n$ is shown in (c).}}
\label{fig:consistency}
\end{figure}

Figure \ref{fig:varym} shows the average error over $100$ replications for estimating the parameters of the BAPLA model using a range of $m$ basis functions. Figure \ref{normal} shows the RMSE and MSE$_f$ for a BAPLA model with $f_i(t/n)$ specified as a scaled normal pdf, where the dashed line at $m=10$ is indicative of the scenario conducted in our earlier simulation study. Whilst we can see that the average RMSE of the interaction matrix begins to plateau (and actually increases for small $n$) after $m=10$, the average MSE$_f$ continues to decrease (for larger $n$) until $m=20$ before plateauing. Therefore, in this particular scenario, one might conclude that a choice of $m=10$ is sufficient for accurate estimation of the interaction matrix, though selecting a larger $m$ could indeed help improve estimates of $f_i(t/n)$. In Figure \ref{gamma}, we give results for the BAPLA model when $f_i(t/n)$ is specified as a scaled gamma pdf. In this scenario, it appears that a choice of $m=5$ would be sufficient for accurate estimation of the interaction matrix as evidenced by the RMSE plot. Moreover, unlike the previous example, the plots show that selecting $m>10$ basis functions would not necessarily improve estimates of the non-stationary firing rate. These results align with our earlier findings from the simulation studies, where we observed that the BAPLA model with $m=10$ basis functions was better able to recover the shape of the gamma pdf compared to that of the normal pdf. Based on these results, we suggest that the original choice of $m=10$ basis functions is indeed sufficient for accurate estimation of the interaction matrix. Moreover, we select $m^*=10$ and $m^* = 20$ as the optimal number of basis functions for the gamma and normal models, respectively. Figures \ref{fig:con_a} and \ref{fig:con_b} show estimates of $f_i(t/n)$ for the gamma and normal models averaged across $100$ replications (solid black line) with empirical $95\%$ confidence bands (dashed lines) and the ground truth in blue. In Figure \ref{fig:con_c} we plot MSE$_f(m^*)$ as a function of $n$, highlighting the somewhat consistent estimation of the non-stationary firing rate as evidenced by the fact that MSE$_f(m^*)\rightarrow 0$ as $n\rightarrow\infty$. Further details of the sensitivity analysis can be found in the Supplementary Material, where additional simulation results are also presented. 

\section{Identifying Neural Connectivity}
\subsection{Data Description and Preprocessing}
 
In this section, we use our proposed BAPLA model to analyse a dataset from the DANDI (Distributed Archives for Neurophysiology Data Integration) archive. DANDI is the BRAIN Initiative\footnote{\url{https://braininitiative.nih.gov/}} supported data archive for publishing and sharing neurophysiology data including electrophysiology, optophysiology and behaviour time series, from various neuroscience experiments. 

Here, we consider a spike train dataset from \cite{steinmetz2019distributed} where Neuropixels probes 
were used to record from a population of neurons in various regions of the mouse brain. In this experiment, neural activity was recorded over a series of experimental trials, while mice performed a visual discrimination task. On each trial, visual stimuli were presented on either the left side, right side, both sides or neither side of a screen at varying contrasts. Mice could earn a water reward by turning a wheel with their forepaws to indicate which side had highest contrast, i.e., the side in which the visual stimulus was most prominent. In the event that neither stimulus was present, mice could earn a reward by keeping the wheel still for $1.5$ seconds. Furthermore, if both the left and the right stimulus had equal non-zero contrast, then the mice earned rewards randomly for turning the wheel left or right.

During the experiment, Neuropixel probes were used to record activity from various regions of the mouse brain. 
At most three probes were inserted at a time, in the left hemisphere, which gave rise to simultaneous recordings from hundreds of neurons in multiple brain regions during each recording session. In total there were $92$ probe insertions over $39$ sessions in $10$ mice. To illustrate the use of the BAPLA model to identify neural interactions, we will consider data from a single mouse during a single recording session, corresponding to the ``sub-Hench\textunderscore ses-20170617T120000.nwb" file on the DANDI archive.

The experiment in \cite{steinmetz2019distributed} was designed to assess the distribution of neurons encoding vision, choice, action, and behavioural engagement across the mouse brain. Here, we attempt to build on their analysis, and construct networks of neuronal interactions for data aligned to either stimulus onset (encoding vision) or  wheel movement (encoding action). In our analysis, we use spiking data from experimental trials in which visual stimuli were presented with the highest contrast. More specifically, we construct four artificial datasets by aligning the spike times either to stimulus onset or wheel movement, and then further split the data according to `left trials' (where the stimulus appeared on the left side of the screen) or `right trials' (where stimulus appeared on the right side). For each dataset, the raw spikes are downsampled with a 1 ms bin size, ensuring that at most one spike can occur per time bin, resulting in a binary time series representation of the neural data. Our goal is to apply the BAPLA model to each dataset individually, in an attempt to visualise neural connectivity in these differing experimental conditions. 

\begin{figure}[t]
\centering
\begin{subfigure}[b]{0.49\textwidth}
    \includegraphics[width =0.9\textwidth]{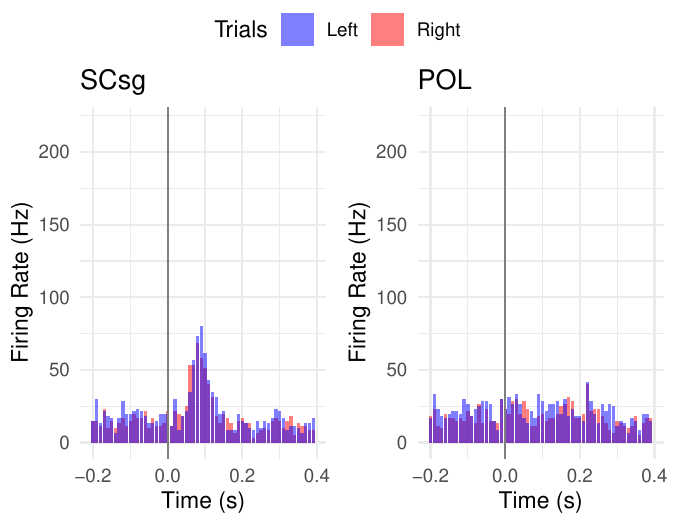}
    \caption{}
    \label{fig:plot_stim}
    \end{subfigure}
    \hfill
    \begin{subfigure}[b]{0.49\textwidth}
    \includegraphics[width =0.9\textwidth]{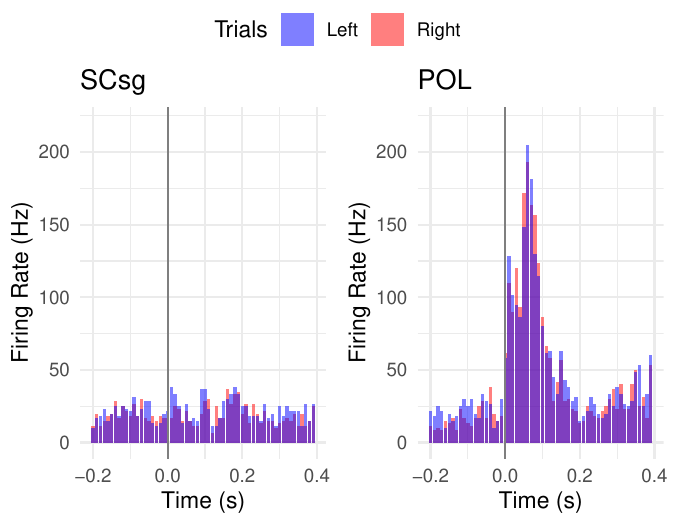}
    \caption{}
     \label{fig:plot_wheel}
    \end{subfigure}
\caption{\small{Firing rate (PSTH plot) aligned to (a) stimulus onset and (b) wheel movement for example neurons in the SCsg and POL regions of the mouse brain.}}
\label{fig:psth}
\end{figure}

Figure \ref{fig:psth} shows firing rates (PSTH plots) aligned to $(a)$ stimulus onset and $(b)$ wheel movement for example neurons in the superior colliculus, superficial grey layer (SCsg) and  the posterior limiting nucleus of the thalamus (POL) regions of the mouse brain. 
Left trials are in blue, while right trials are in red. In these plots, we observe an increase in the firing rate of the SCsg neuron following onset of the visual stimulus, but no response following wheel turns, while the converse is true for the POL neuron. Throughout, we will refer to neurons who do not respond to stimulus onset or initiation of the wheel movement as `stationary', and those whose firing rate increases as `non-stationary'.

Initial exploratory analysis revealed that individual neurons do not necessarily fire in every single experimental trial. Therefore, to avoid subsequent issues with the estimation procedure, we choose to remove these trials (in which there are no spiking events) from our analyses. Moreover, to enable a fair comparison between the left and right trials for a specific experimental condition, we analyse neural data from the most active, i.e., highest firing, $l=60$ trials, yielding an effective sample size of length 36 seconds ($60$ trials each of length $600$ms). Finally, neurons are only selected for analysis if at least $10$ spikes are observed per trial on average. As we saw in Figure \ref{fig:psth}, individual neurons may or may not respond at different stages in the experimental procedure. However, since neurons are selected for analysis based on their firing rates, this means that some neurons used in the analysis of the wheel movement data are excluded from the analysis of the stimulus onset data, and vice versa. Therefore, to enable a fair comparison between the groups, all neurons are plotted in our estimated networks of neural connectivity (Figure \ref{fig:graphs}) where white nodes represent neurons excluded from the analysis of a particular dataset. Overall, we consider a subset of $82$ neurons from $11$ different brain regions that exhibit sufficiently high average firing rates in the period of interest, i.e., $-0.2s$ before and $0.4$s after onset of the visual stimulus or initiation of the wheel movement.

To estimate the neural connectivity among our population of neurons, we consider the following adaptation to the regularised maximum likelihood estimator \eqref{obj_unc}  in which we also incorporate the trial-like structure of the data. Letting $L_l(\theta, \phi_i(t/n), \lambda)$ be the trial-specific loss according to \eqref{obj_c}, we obtain parameter estimates by maximising the joint likelihood according to 
\begin{align}
 \hat{\theta}=\arg\max_{\boldsymbol{\theta}} \sum_{l=1}^{\#trials} L_l(\theta, \tilde{\phi}_i(t/n), \lambda),
   \label{eq:trial_obj}
\end{align}
assuming independence across trials.
As before, we solve the above optimisation problem \eqref{eq:trial_obj} using a coordinate descent algorithm and select the regularisation parameter using a BIC-type criterion. The update steps, accounting for the trial-like structure of the data, are given in the supplementary material. Moreover, we estimate the non-stationary firing rate using $m=6$ basis functions.

\subsection{Graphical Representation of Neural Connectivity}

\begin{figure}[t]
\centering
\begin{subfigure}[b]{\textwidth}
    \includegraphics[width =0.9\textwidth]{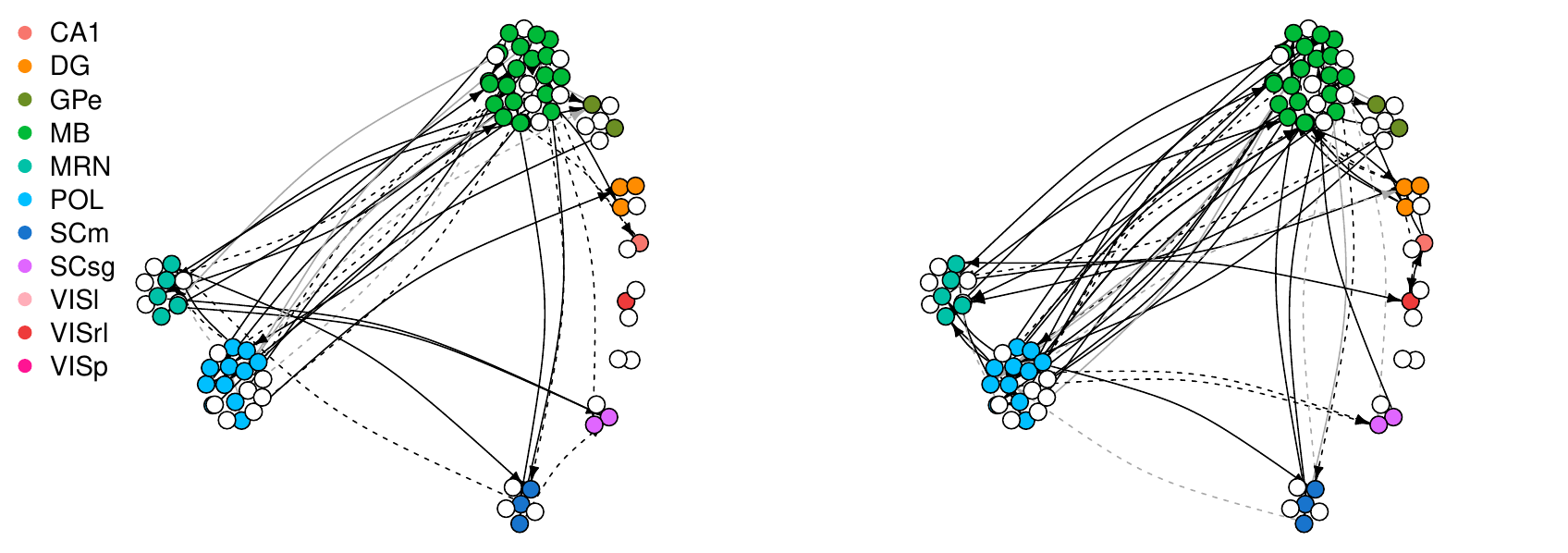}
    \caption{}
    \label{fig:interaction_stim}
    \end{subfigure}
    \hfill
    \begin{subfigure}[b]{\textwidth}
    \includegraphics[width =0.9\textwidth]{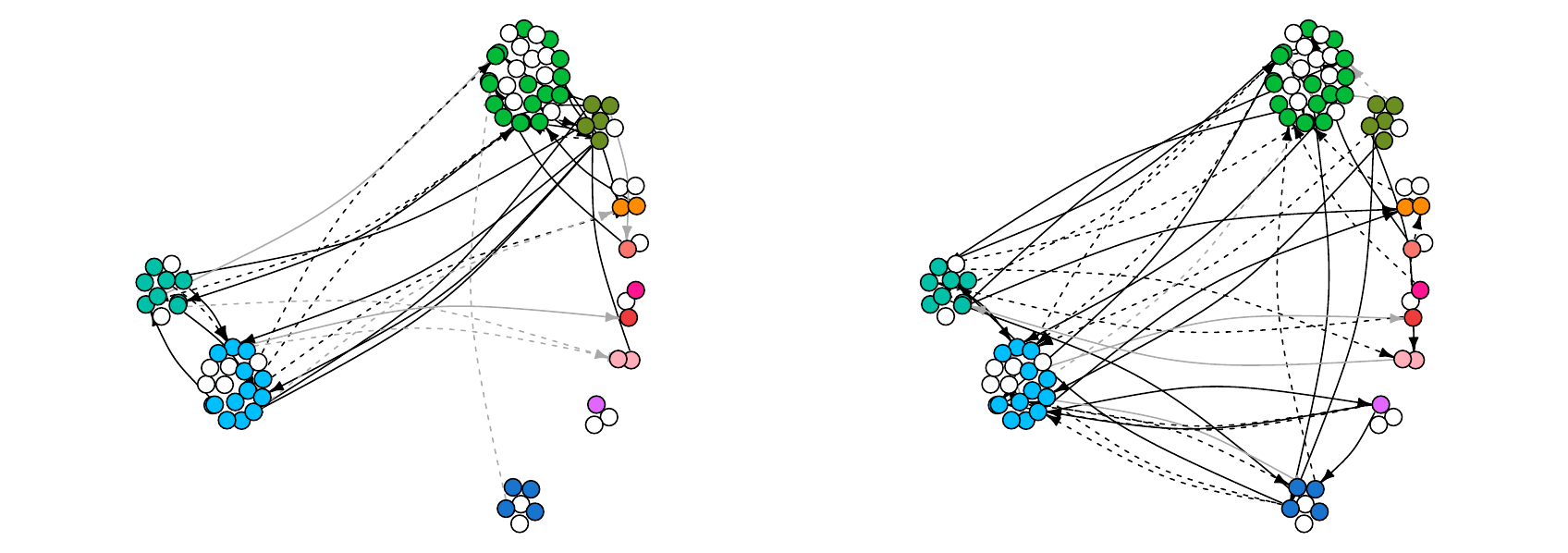}
    \caption{}
     \label{fig:interaction_wheel}
    \end{subfigure}
\caption{\small{Estimated networks for left and right trials aligned to (a) stimulus onset and (b) wheel movement. Solid lines represent excitatory edges and dashed lines indicate inhibitory edges. White nodes represent neurons omitted from the analysis due to insufficient firing rates and grey edges denote entries in the interaction matrix deemed insignificant by our inference procedure.}}
\label{fig:graphs}
\end{figure}

Graphical representations of the estimated interaction matrices are shown in Figure \ref{fig:graphs}. In these graphs, each node represents a neuron (colour coded by brain region\footnote{\url{https://atlas.brain-map.org/}}) and a directed edge from node $i$ to node $j$ represents neuron $i$'s influence upon neuron $j$, accounting for the behaviour and activity of all the other neurons. White nodes represent the neurons which are omitted from the analysis due to low firing rates. A solid line represents an excitatory influence (i.e., $\Gamma_{ij}>0$) and a dashed line indicates an inhibitory influence (i.e., $\Gamma_{ij}<0).$  Absence of an edge between two neurons means there is no interaction detected via the BAPLA model.

Networks of the detected interactions aligned to stimulus onset and wheel movement are presented in Figures \ref{fig:interaction_stim} and \ref{fig:interaction_wheel}, respectively. For the stimulus onset data, more edges are detected for the right trial dataset (68) compared to the left trial dataset (49). This is also true for the wheel movement dataset, where we observed 51 edges for right trials compared to only 38 edges for left trials. 
In both cases, the BAPLA model detects more excitatory edges than inhibitory ones. However, this is somewhat as expected since it is generally harder to detect inhibitory edges, especially if neurons exhibit low firing rates because further inhibition is limited by a floor at zero \citep{zhao20121}. 
In general, our BAPLA model identifies more between region interactions compared to those in the same region (see Table \ref{tab:summary_stats} for further details).

For the stimulus onset data, we found that $44\%$ of the total number of interactions are between neurons in the midbrain (MB) and the posterior limiting nucleus of the thalamus (POL). Interactions between these brain regions are part of a wider circuit known to facilitate crucial brain functions like sensory processing and motor control \citep{inagaki2022midbrain}. While POL neurons also interact with neurons in other midbrain nuclei such as MRN and SCm, we found that neurons in the thalamus (POL) do not interact much with each other; an observation also highlighted by \cite{jager2021dual}. In particular, we identified only 2 edges in rightward trials and 3 edges in leftward trials among POL neurons. Interactions among midbrain nuclei (MB, MRN, SCm and SCsg) account for $\approx 36\%$ of the total number of interactions, and these regions also interact with neurons in the basal ganglia (GPe).
Overall, our results for the stimulus onset data align with the findings in \cite{steinmetz2019distributed} where it is highlighted by neuroscientists that neurons encoding vision were found in a pathway comprising primarily of visual areas, such as the thalmus (POL) and superficial superior colliculus (SCs), and other structures such as the basal ganglia (GPe) and several midbrain nuceli (SCm, MRN).

\begin{table}[t]
    \centering
    \footnotesize
    \begin{tabular}{ccccccc}
        \multirow[b]{2}{*}{Dataset} & \multirow[b]{2}{*}{Trial} 
        & \multirow[b]{2}{*}{No. Edges} & \multicolumn{2}{c}{\text { Interaction Type $(\%)$}} & \multicolumn{2}{c}{\text { Region Type $(\%)$}}  \\
        
        & & & \text {Excitatory} & \text { Inhibitory } & \text {Within} & \text {Between} \\ 
        \\
        Stimulus Onset & Left & 49 & 61.22 & 38.78 & 24.49 & 75.51 \\
        &               Right & 68 & 69.12 & 30.88 & 14.71 & 85.29 \\
        \\
        Wheel Movement & Left & 38 & 65.79 & 34.21 & 15.79 & 84.21 \\
        &               Right & 51 & 58.82 & 41.18 & 19.61 & 80.39 
    \end{tabular}
    \caption{Summary statistics for estimated interaction matrices.}
    \label{tab:summary_stats}
\end{table}

For the wheel movement dataset, we found that $75\%$ of the total number of interactions are between neurons in the midbrain nuclei (MB, MRN, SCm and SCsg), the posterior limiting nucleus of the thalamus (POL) and the basal ganglia (GPe). Interactions among these particular brain regions are central to the function of the so-called cortico-basal ganglia-thalamo-cortical (CBGTC) loop. The CBGTC is one of the fundamental network motifs in the brain, and revealing its structural and functional organisation is critical to understanding cognition, sensorimotor behaviour and the natural history of many neurological and neuropsychiatric disorders \citep{foster2021mouse}. {In general, neural connectivity appears to be more widespread across the different brain regions for the wheel movement dataset compared to the stimulus onset data, particularly for the right trials. Our results therefore corroborate the findings of \cite{steinmetz2019distributed} who highlight that neurons encoding action were spread throughout all recorded regions, while neurons encoding visual stimuli were more constrained to a particular pathway.}

We construct confidence intervals for entries in the interaction matrix using the desparsified estimator outlined in Section \ref{sec:inference}. If a given interval contains zero, the associated entry in the interaction matrix is deemed insignificant and is consequently set to zero, consistent with the sparsity-inducing nature of our regularised estimator. {Using this approach, we find that $8 \ (9)$ edges in the left trials and $11 \ (10)$ edges in the right trials are deemed insignificant at a $5\%$ significance level for the stimulus onset (wheel movement) dataset. These so-called insignificant edges are highlighted in grey in Figure \ref{fig:graphs}.}

\subsection{Accounting for Non-Stationary Firing Rates}

\begin{figure}[t]
    \centering
    \includegraphics[width=0.9\textwidth]{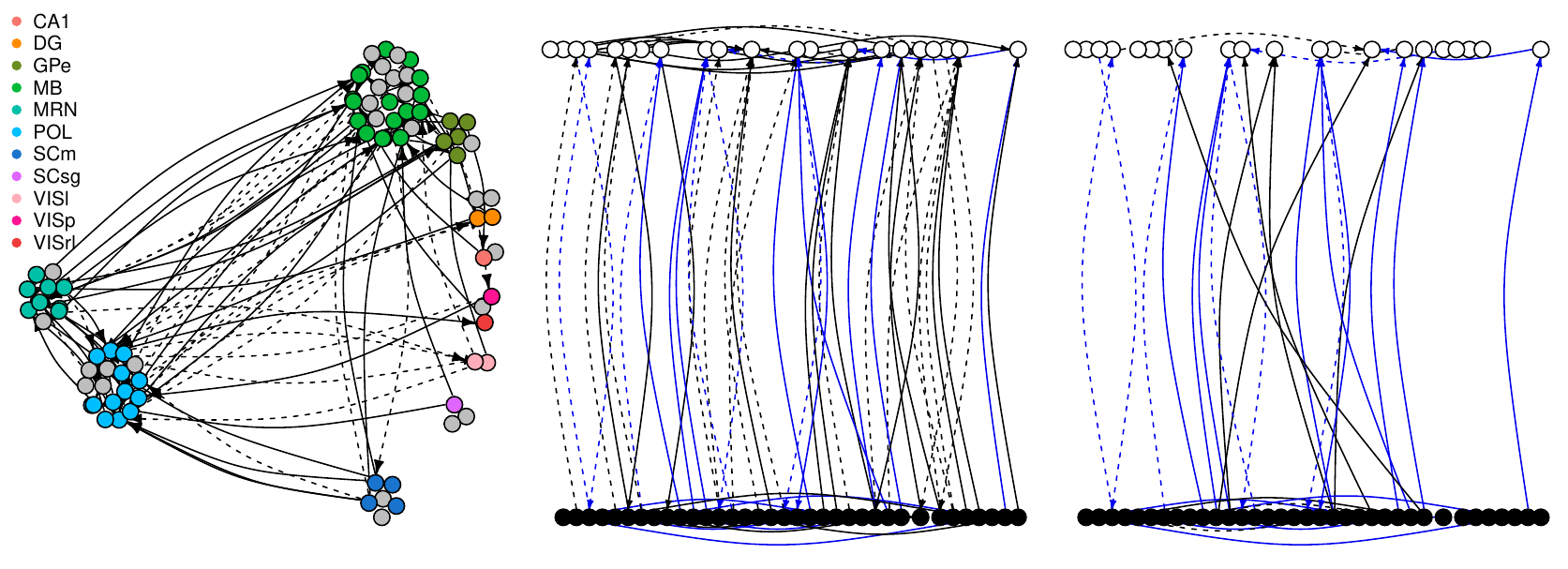}
    \caption{\small{Estimate of the interaction matrix for left trial data in the wheel movement dataset with standard logstic model for all regions (left) and split into stationary and non-stationary neurons (middle). Non-stationary neurons are depicted by white nodes and stationary neurons are in black. Figure on right shows estimate of the interaction matrix obtained using the BAPLA method. Common edges between the BAPLA model and logistic model are in blue. }}
    \label{fig:compar_f}
\end{figure}

In this section, we assess to what extent accounting for non-stationary firing rates in our BAPLA model impacts the estimation of the interaction matrix. Figure \ref{fig:compar_f} shows the estimated interaction matrix under the standard logstic model (which assumes stationary firing rates) for left trials in the wheel movement dataset (left). 
Generally, more edges are detected under the logistic model (80) compared to our proposed BAPLA method (38), perhaps suggesting that neurons with non-stationary firing rates are more likely to interact; either with each other, or with other stationary neurons. To explore this possibility, we sort the spike train data into two groups, and classify each neuron as either stationary or non-stationary as evidenced by their firing rate. 

Figure \ref{fig:compar_f} shows estimates of the interaction matrix under the logistic model (middle) compared to that obtained using our BAPLA method (right). In these plots, the black and white nodes represent stationary and non-stationary neurons, respectively, and edges common to both estimates are shown in blue. Displaying the networks in this way, we can see that accounting for non-stationary firing rates reduces the number of edges, both within the group of non-stationary neurons and between other stationary neurons. In particular, 19 edges are detected between non-stationary neurons under the logistic model, compared to only 4 for the BAPLA model. Additionally, the number of edges between the groups reduces from 44 to 21 when we use our proposed BAPLA method. Overall,
our results suggest that explicit modelling of the firing rate is essential to reduce the number of spurious connections in the estimated interaction matrix.

\section{Discussion}
To conclude, we have proposed a Bernoulli autoregressive partially linear additive model to identify excitatory and inhibitory interactions among a population of non-stationary neural spike trains.  Our results demonstrate that our estimation procedure works well in a variety of scenarios, and that explicit modelling of the non-stationary firing rate is essential for accurate recovery of the interaction matrix.  
We also outline an inference procedure that can be used to construct confidence intervals for our estimates of neural connectivity.
Furthermore, we have demonstrated the usefulness of our model in practice using a real dataset of scientific interest, where we study interactions between neural processes using spike train data from \cite{steinmetz2019distributed}. 

This paper extends the current literature for identifying neural connectivity in two ways: 1) by accounting for non-stationary firing rates in the estimation procedure, and 2) by providing confidence intervals for entries in the interaction matrix. However, there are many possible and interesting avenues for future work. From an application point of view, extensions of the BAPLA model might include the addition of some categorical covariates or behavioural variables, specific to the neuroscience experiment of interest. Methodologically, our method could be extended to include an additional penalty term in the regularised maximum likelihood estimator \eqref{obj_c} for automatic selection of the number of basis functions, or to enforce smoothness conditions on $f_i(t/n)$. For the purposes of this work, we argue that the benefits of including an additional penalty term do not outweigh its complications, nor its increase in computational complexity. However, we believe that in some scenarios and alternative applications this development could be particularly advantageous. Finally, it remains to provide a full theoretical analysis for our adaptation of the desparsified estimator, to construct confidence intervals for entries in the interaction matrix. While our simulation results demonstrate empirical support for the validity of our proposed inference approach, future work could involve tracking the asymptotic properties of the desparsified estimator in the presence of non-linear components. Perhaps the work of \cite{gueuning2016confidence} or \cite{drikvandi2025high} looking at a combination of $\ell_2$ and $\ell_1$ penalisation might provide a reasonable starting point for such analysis.

\section*{Data} 
The data that support the findings of this study are openly available on the DANDI Archive (RRID:SCR$\_017571$) at \url{10.48324/dandi.000017/0.240329.1926} \citep{steinmetz2024distributed}.

\section*{Code}
The code used to implement the BAPLA model is available at \url{https://github.com/pinkney19/BAPLA}.

\bibliographystyle{apalike}
\bibliography{refs}

\end{document}